\documentclass{article}

\usepackage{arxiv}

\usepackage[utf8]{inputenc}
\usepackage{graphicx}
\graphicspath{{images/}{figures/}}
\usepackage{amsmath}
\usepackage{amssymb}
\usepackage{booktabs}
\usepackage{multirow}  
\usepackage{enumitem}  
\usepackage{longtable}  
\usepackage{microtype}  
\usepackage[T1]{fontenc}
\usepackage{float}
\usepackage{lmodern}
\usepackage{titlesec}
\usepackage{url}  
\usepackage{makecell}
\usepackage{booktabs}
\titlespacing*{\paragraph}{0pt}{1ex plus 0.2ex}{1em}
\usepackage{multicol}

\usepackage{url}  

\usepackage[colorlinks=true,linkcolor=blue,citecolor=blue,urlcolor=blue]{hyperref} 

\begin{document}
\sloppy

\title{State-of-the-art Small Language Coder Model: Mify-Coder}
 \author{Infosys AI Research | Mify Team\thanks{A detailed contributor list can be found in the appendix of this paper.}}

\maketitle

\begin{abstract}
We present Mify-Coder, a 2.5B-parameter code model trained on 4.2T tokens using a compute-optimal strategy on Mify-2.5B\footnote{\scriptsize \mbox{\url{https://blogs.infosys.com/topaz/nvidia/optimizing-enterprise-ai-navigating-llm-challenges-and-rise-of-slms.html}}} foundation model. Mify-Coder achieves comparable accuracy and safety while outperforming significantly larger baseline models on standard coding and function-calling benchmarks, demonstrating that compact models can match frontier-grade models in code generation and agent-driven workflows. Our training pipeline combines high-quality curated sources with synthetic data generated through agentically designed prompts, refined iteratively using enterprise-grade evaluation datasets. LLM-based quality filtering further enhances data density, enabling frugal yet effective training. Through disciplined exploration of CPT–SFT objectives, data mixtures, and sampling dynamics, we deliver frontier-grade code intelligence within a single continuous training trajectory. Empirical evidence shows that principled data and compute discipline allow smaller models to achieve competitive accuracy, efficiency, and safety compliance. Quantized variants of Mify-Coder enable deployment on standard desktop environments without demanding specialized hardware. 
\end{abstract}

\section{Introduction}
Large language models (LLMs) have emerged as transformative tools in software engineering, enabling automated code generation, intelligent debugging, and agent-driven workflows. These capabilities promise significant productivity gains and novel development paradigms. However, the dominant trajectory of progress has been scale-centric: state-of-the-art code models often comprise tens or hundreds of billions of parameters. While these architectures deliver impressive performance across programming tasks and reasoning benchmarks, they impose substantial computational and financial burdens. Training a 30B-parameter model on trillion-scale datasets, for instance, can require over a million GPU hours, depending on hardware and optimization strategies, rendering such solutions inaccessible to most enterprises and research institutions.
 
Efforts to mitigate these constraints have explored compute-efficient strategies, such as smaller architectures. While these approaches reduce resource demands, achieving enterprise-grade intelligence further depends on factors, such as data quality and safety alignment. Prior work has largely addressed these dimensions in isolation; however, integrated strategies that unify compute efficiency, robust data curation, and safety alignment within a single continuous training pipeline remain underexplored.
 
This research addresses above limitations by investigating whether smaller models, when trained with rigorously engineered strategies, can achieve frontier-grade performance without incurring prohibitive resource requirements. We introduce \textbf{Mify-Coder}, a 2.5B-parameter code model designed to deliver competitive accuracy and enterprise-ready safety at a fraction of the cost of scale-first paradigms. Our approach integrates continued pretraining (CPT) and supervised fine-tuning (SFT) within a compute-optimal trajectory, supported by disciplined objective scheduling, curated data pipelines, and alignment mechanisms embedded from the outset.
 
\subsection*{Core Design Principles}
\begin{enumerate}
    \item \textbf{Compute-Optimal Training:} A precisely engineered CPT--SFT recipe executed on a compact cluster of B200 and H100 GPUs within three months.
    \item \textbf{Robust Data Pipeline:} Integration of vetted open-source repositories with synthetic data generated from agentically curated prompts and LLM-based quality filtering.
    \item \textbf{Enterprise-Grade Safety:} Alignment with Responsible AI principles to ensure harmlessness, helpfulness, and secure code generation.
    \item \textbf{Deployment Practicality:} Quantized variants enabling inference on standard CPU environments without GPUs.
\end{enumerate}
 
By demonstrating that Mify-Coder rivals instruction-tuned models such as \textit{DeepSeek-Coder} and \textit{Qwen2.5-Coder} on widely used benchmarks such as \textit{HumanEval}, \textit{MBPP}, and \textit{BFCL} while offering superior deployment flexibility, this work provides evidence that capability is not exclusively a function of scale. Through compute discipline, high-quality data, and robust alignment strategies, compact models can deliver sustainable, high-performing solutions for modern software development. Mify-Coder model weights will be open-sourced to accelerate research and real-world adoption.
\section{Related Work}
Language models have progressed from statistical n-gram approaches to neural architectures and Transformer-based designs, enabling large-scale training and general-purpose reasoning. While large models dominate benchmarks, smaller models have gained attention for their efficiency and adaptability in resource-constrained environments, making them suitable for enterprise and specialized applications.

Several open-source models target code-related tasks with smaller parameter footprints. PolyCoder (2.7B)  \cite{xu2022polycoder}  focuses on C code, SantaCoder~\cite{allal2023santacoder} (1.1B) supports multiple languages using fill-in-the-middle objectives, and TinyStarCoderPy\footnote{\url{https://huggingface.co/bigcode/tiny_starcoder_py}} (164M) offers a lightweight Python variant. CodeT5-small \cite{wang2021codet5} employs an encoder–decoder architecture for code summarization and translation, while Qwen2.5-Coder introduces smaller variants optimized for multi-language coding and long-context reasoning. These models reflect a growing trend toward balancing efficiency with capability, though most remain limited in scale and enterprise integration.

Open-source initiatives such as StarCoder2 \cite{lozhkov2024starcoder2}, WaveCoder \cite{yu2023wavecoder}, and OctoPack \cite{muennighoff2023octopack} demonstrate the effectiveness of instruction tuning and commit-based datasets for improving code generation. However, these efforts primarily target larger models, leaving a gap for enterprise-grade solutions that combine efficiency, security, and accuracy. Building on these trends, our work addresses this gap by adapting a 2.5B-parameter enterprise model into a specialized coding system. We leverage curated enterprise code and high-quality synthetic data augmentation to achieve competitive results on standard benchmarks while maintaining resource efficiency.

Recent work highlights that data quality is as critical as scale for code LLM performance. Arctic-SnowCoder \cite{wei2024arctic} employs a three-phase strategy, combining large-scale raw code, curated subsets, and synthetic data augmentation, demonstrating strong gains with data-efficient pretraining. Similarly, instruction tuning and synthetic data generation, as seen in Magicoder \cite{wei2024magicoder} and OpenCodeInstruct \cite{ahmad2025opencodeinstruct}, have become essential for improving robustness and security in code generation. These trends reinforce our approach of leveraging curated enterprise code and high-quality synthetic augmentation to achieve competitive results while maintaining resource efficiency.
\section{Data}

\subsection{Data Sources and Composition}

As part of our core methodology, we adhered to four foundational principles in our dataset sourcing strategy: Compliance, Diversity, Density, and Quality. To design our data acquisition strategy, we analyzed approaches adopted by coder models, including SantaCoder, DeepSeek-Coder~\cite{guo2024deepseek}, and Qwen2.5-Coder~\cite{qwen2.5}. We use combination of their approaches for data sourcing, quality checks and data mixing strategy. To reduce the data processing time and budgets, we relied on existing datasets instead of curating from raw sources where possible.

All data collection adhered to organizational security and compliance protocols. Candidate datasets underwent rigorous IP and Responsible AI reviews, with comprehensive license and provenance checks to exclude restricted datasets and those generated from models which restrict the usage of its data for training.

The final training corpus reflects a carefully balanced composition across three primary modalities, optimized through extensive ablation studies to support comprehensive code generation, mathematical reasoning, and natural language understanding capabilities. The optimal composition employs a 78:12:10 ratio (code:text:math), determined through systematic ablation studies detailed in Section~\ref{sec:data-ablation}.

\noindent\textbf{Source Code Data}

For pre-training, for code data, primarily we relied on the Stack-v2 dataset~\cite{allal2023santacoder} via Software Heritage; the dataset does not include copyleft and commercial license code. Stack-v2 data pipeline ensures removal of non-code data and headers. Additionally, it filters configuration files like HTML, JSON, YAML, Markdown to minimize the files. The dataset encompasses diverse programming languages with a focus on six primary languages: Python, Java, C\#, C++, C, and JavaScript, along with other languages such as SQL, HTML, XML, and Markdown.

To include high quality tokens, we also reformatted instruction tuning and reasoning datasets. The source code component comprises 78\% of the total training tokens.

\noindent\textbf{Mathematical Data}

Math is a fundamental capability for code-focused models because many programming tasks inherently involve mathematical reasoning, such as algorithm design, numerical computation, and data structure manipulation.
We relied on sourcing Math data from existing filtered sources instead of extracting from common crawl dumps.

As demonstrated by prior work such as Phi-4-Mini~\cite{abouelenin2025phi4mini}, Llama3\footnote{\url{https://ai.meta.com/blog/meta-llama-3/}}, and Qwen3~\cite{yang2025qwen3}, we allocated approximately 15\% of the total token budget to math data during the initial Code Pre-Training (CPT) stage, and 10\% of which for high-quality math data. To increase high quality coverage, the dataset was augmented with additional SFT and reasoning math datasets. Our total mathematical data constitutes 10\% of the training corpus.

\noindent\textbf{Text Data}

Text data plays a crucial role in enabling code models to understand natural language instructions, technical documentation, and contextual reasoning. We sourced our text corpora from high-quality, openly available datasets, code-related documentation, scientific literature, GitHub repositories, and curated sources such as Wikipedia. 

We prioritized domain-specific STEM content to provide essential context for code understanding, algorithm explanation, and technical documentation comprehension. This selection ensures the model can effectively bridge natural language descriptions with code implementations. Text data constitutes 12\% of the training corpus, complementing the code and mathematical components to support comprehensive language-code understanding.

\noindent\textbf{Instruction Data}

To support advanced coding capabilities, we prioritized tasks such as bug fixing, completion, generation, reasoning, function calling, and mathematical reasoning. Our sourcing strategy focused on English language content across the six primary programming languages, evaluating open-source Code SFT and reasoning datasets along with Math SFT datasets through rigorous license and provenance checks.

After reviewing dataset cards, documentation, and related papers, we shortlisted permissible datasets through multiple review stages as listed below:

\begin{table}[htbp]
\centering
\footnotesize
\caption{Instruction Data Filtering Stages.}
\begin{tabular}{clp{5.5cm}c}
\toprule
\textbf{Stage} & \textbf{Title} & \textbf{Description} & \textbf{Retention} \\
\midrule
1 & License and Provenance Validation & Initial screening based on license compliance, dataset provenance, exclusion of closed-source model outputs, and size constraints. & 50\% \\
\midrule
2 & Quality-Based Filtering & Application of filtering mechanisms leveraging trained model scores and secondary categorization. & 18\% \\
\midrule
3 & Dataset Impact Analysis & Ablation studies to evaluate individual dataset contribution to model performance (Section~\ref{sec:sft}). & 3\% \\
\midrule
4 & Decontamination & See Section~\ref{sec:decontamination}. Drop was minimal as datasets were pre-reviewed. & 3\% \\
\midrule
5 & Metadata-Driven Annotation & Bucketing for usage across stages (Section~\ref{sec:instruction-quality}). & Bucketed \\
\bottomrule
\end{tabular}
\label{tab:instruction-stages}
\end{table}

For long-term scalability, datasets were bucketed by Natural language type, Token count and Programming language coverage. This structured organization enables efficient folder selection for training and facilitates future expansion into additional domains.

The final curated collection comprises diverse tasks including generation, completion, translation, summarization, defect detection, SQL generation, docstring generation, test case synthesis, mathematical reasoning, chain-of-thought reasoning, and tool-calling examples.

\subsection{Data Cleaning \& Preprocessing}

All datasets were processed through distributed preprocessing pipelines built on frameworks such as Ray, Dask, and DAFT to ensure scalability and efficiency. Dedicated pipelines were designed for different data types, code, mathematics, and text, with tailored transformations to accommodate dataset-specific characteristics and training objectives.

Code datasets underwent multi-stage cleaning processes to remove non-value headers, different tags, boilerplate text in headers, which ensured meaningful code was retained for the given context length.

Consistent with prior approaches proposed in works such as DeepSeek-Coder~\cite{guo2024deepseek}, Qwen2.5-Coder~\cite{qwen2.5}, and SantaCoder~\cite{allal2023santacoder}, we employed a dependency-aware topological sorting strategy across all files within each repository by parsing inter-file relationships. This design ensures that the model captures project-level context in a structured and coherent manner. We did not include any repos which had only 1 file.

Analysis revealed that most math data originated from similar sources, with minimal variation beyond extraction and formatting. We identified following challenges: Limited diversity due to heavy reliance on common sources of data, and high duplication requiring aggressive deduplication strategies.

From an initial pool of approximately 30 mathematical datasets, we removed approximately 70\% tokens by doing provenance analysis. After multiple experiments on fuzzy deduplication using MinHashLSH, we finalized MinHash with shingle size of 5, 110 permutations, 10 LSH bands, 11 minhashes per band with a Jaccard similarity threshold of 0.75, resulting in a drop of further 20\% tokens.

Since our Text data sources and above identified Math data sources are derived from common web sources, we employed URL-based filtering and removed all Math related sources from Text data sources, which reduced 3\% of token overlap. The text dataset underwent domain-specific filtering using NVIDIA Domain Classifier\footnote{\url{https://huggingface.co/nvidia/domain-classifier}} to ensure we can select tokens which are close to related domains. Document-type bucketing (Academic Writing, Knowledge Article, Documentation) using EAI-Distill models\footnote{\url{https://huggingface.co/QuantFactory/eai-distill-0.5b-GGUF}} captured structural and stylistic variations critical for specialized learning objectives.

\subsection{Filtering and Quality Control}

We implemented a comprehensive, multi-stage filtering and quality control pipeline applied uniformly across all data categories.

\subsubsection{Language Identification and Rule-Based Filtering}

Based on initial ablations on the text dataset, we found that dataset contains non-English characters, we employed confidence-based filtering discarding samples with low-confidence language assignments. Binary files, auto-generated templates, and malformed content were excluded through rule-based filtering with strict file size and extension allowlists.

\subsubsection{Code Quality Assessment}

High-quality code density is critical for effective model training, as supported by recent research. Although numerous approaches have been proposed to assess data quality, no single method provides a definitive measure of overall quality. We developed a high-density, signal-driven framework to categorize code quality in the stack\_v2 dataset into three buckets: High, Medium, and Low.

\paragraph{Signal Framework}

The quality assessment framework encompasses multiple dimensions of code quality. Post analyzing 50+ quality signals spanning WaveCoder~\cite{yu2023wavecoder} and SantaCoder~\cite{allal2023santacoder}, using rigorous distributional analysis including histograms, coverage checks, and applicability reviews along with manual analysis of data, we zeroed on:

\begin{enumerate}
\item Textual and structural statistics capture content composition through alphanumeric ratio, fraction of characters in quoted strings, encoded data \%, duplicate grams, lines per function, whitespace ratio, \% of `print' and `todo' kind of statements and char token ratio.
\item Comments and documentation quality are evaluated through comment-to-code ratio, docstring density, and in-line explanation frequency.
\item Model-based signals provide sophisticated quality assessments through ArmoRM~\cite{wang2021codet5}, which evaluates code readability, explainability, and complexity, and StackEduClassifier~\cite{abouelenin2025phi4mini}, which provides language-conditioned scores across 15 programming languages reflecting educational clarity and pedagogical structure.
\end{enumerate}

\subsubsection{Classifier-Based Quality Filtering}

Neural classifier models were applied to assign quality scores to both textual and code artifacts, enabling threshold-based selection of high-quality samples. For text data, the NVIDIA Quality Classifier identified content suitable for pre-training, while EAI-Distill models enforced strong correctness, completeness, and clean extraction signals. For mathematical content, the FineMath classifier enabled quality stratification and bucketed subsets for different training stages.

\subsubsection{Near-Duplicate Removal and Code Validation}

Both n-gram-based and semantic similarity metrics were utilized to identify and remove near-duplicate content, preserving data diversity while eliminating redundancy. For code artifacts, Abstract Syntax Tree (AST) parsing and selective execution checks ensured syntactic validity and plausible runtime behavior, filtering out malformed or non-executable code.

\subsubsection{Instruction Data Quality Assessment}
\label{sec:instruction-quality}

A review of instruction data applied an LLM-as-a-judge workflow for quality assessment (completeness, clarity, coding-standard adherence). We used Qwen2.5-Coder (70B)~\cite{qwen2.5}, Llama3-70B-Instruct, and Claude 3.5 Sonnet\footnote{\url{https://www.anthropic.com/news/claude-3-5-sonnet}} as independent judges to score relevance, formatting consistency, and correctness. This yielded a high-quality subset from the initial instruction pool.

\subsection{Decontamination}
\label{sec:decontamination}

To ensure robust evaluation and prevent data leakage, we adopted an n-gram-based decontamination approach informed by NVIDIA NeMo Curator\footnote{\url{https://docs.nvidia.com/nemo/curator/latest/index.html}} and Xcoder, detecting both exact overlaps and rewritten variants of benchmark content. Optimal parameters were determined through systematic evaluation using standard metrics including precision, recall, and F1-score, achieving perfect recall and high precision. Drop was minimal as the datasets were chosen based on initial detailed reviews.

All sourced and synthetic datasets were systematically decontaminated against comprehensive benchmark suites spanning code generation, reasoning and debugging, function calling, and mathematics. This process ensured training data remained free of benchmark leakage, preserving evaluation integrity.

\subsection{Synthetic Data Generation}
Even after substantial curation of real-world repositories, gaps remain in domain coverage, task diversity, and representation of long-tail programming scenarios. Synthetic data generation therefore serves as a complementary mechanism, enabling controllable expansion of training distributions, targeted capability strengthening, and scalable creation of multi-turn interaction data that cannot be reliably obtained from public sources.

\subsubsection{Generation Strategy}
Our approach integrates seed-grounded generation, seedless generation, and agentic data enrichment:

\begin{itemize}
    \item \textbf{Seed-grounded expansion:} We construct a large set of narrowly defined topics (“seed data”), generated via LLM-driven topic decomposition from broad domains. Each seed includes detailed descriptions, constraints, and few-shot exemplars, ensuring distributional fidelity to real-world tasks. Synthetic instances are then produced per-topic to maximize diversity while avoiding mode collapse.

    \item \textbf{Seedless generation with prompt rotation:} When seed examples are unavailable, we employ structured prompt templates and rotate few-shot exemplars to broaden coverage while maintaining task relevance.

    \item \textbf{Agentic multi-turn generation:} For interaction-heavy tasks, such as coding assistance, diagnostics, and iterative reasoning, we use multi-agent roleplay. Paired agents (e.g., developer $\leftrightarrow$ coding assistant, student $\leftrightarrow$ tutor) generate coherent multi-turn dialogues where follow-up questions and clarifications emerge naturally based on prior turns. We used Qwen2.5-Coder and Llama3 Instruct as primary models for agentic multi-turn generation and for most synthetic data workflows, ensuring alignment with task objectives and generation strategy.
\end{itemize}
\subsubsection{Quality Validation}
All synthetic outputs undergo a multi-stage validation pipeline:

\begin{itemize}
    \item \textbf{Semantic deduplication:} Embedding-based clustering removes near-duplicates across both real and synthetic sources.
    \item \textbf{Format and structural validation:} Automated schema checks enforce consistency with instruction-tuning, code-generation, or conversation formats.
    \item \textbf{Benchmark decontamination:} We filter out instances overlapping with evaluation suites using fuzzy matching and semantic similarity thresholds.
    \item \textbf{LLM-as-judge evaluation:} High-capacity evaluators (Qwen2.5-Coder (70B)~\cite{qwen2.5}, Llama3 Instruct (70B), and Claude 3.5 Sonnet score outputs for relevance, correctness, coherence, and safety.
    \item \textbf{Role-consistency filtering (for multi-turn data):} Additional checks ensure conversational agents maintain their assigned personas and dependencies across turns.
\end{itemize}
\subsubsection{Volume and Scale}

Systematic ablations were conducted to determine optimal synthetic data volumes. The baseline SFT configuration utilized 3B tokens across 375,000 samples. Synthetic-only experiments employed 3B tokens across 250,000 samples. Tool, function calling, and multi-turn dialogue augmentation increased the corpus to 4B tokens across 500,000 samples. Scaling sweeps explored configurations at 8B, 12B, and 15B tokens, corresponding to 1M, 1.5M, and 1.875M samples respectively.

\begin{table}[htbp]
\centering
\caption{Volume of Synthetic Data Ablations.}
\begin{tabular}{lrr}
\toprule
Ablations & Token Budget (in B) & \#Samples \\
\midrule
Baseline SFT & 3 & 375,000 \\
Synthetic-only & 3 & 250,000 \\
Tool/Function \& Multi-turn & 4 & 500,000 \\
Scaling sweeps & 8/12/15 & 1M / 1.5M / 1.875M \\
\bottomrule
\end{tabular}
\end{table}

\section{Training Details} 
\subsection{Training Overview}
\textbf{Mify-2.5B} serves as the base model for our work. Building on this foundation, we developed \textbf{Mify-Coder} by performing two key training phases:

\begin{itemize}
    \item \textbf{Continual Pretraining (CPT):} Establish broad code intelligence via next-token prediction, optionally augmented with Fill-in-the-Middle (FIM) for structural infilling skills.
    \item \textbf{Supervised Fine-Tuning (SFT):} Align model behavior to user-facing tasks such as single-turn coding, multi-turn dialogs, tool/function calling, mathematical reasoning, and safety requirements.
\end{itemize}

To run various pipelines - data processing/annotation pipelines, synthetic data generation and training, we have hosted two separate Nvidia DGX B200 and H100 clusters built with in our datacenter, with Nvidia Infiniband which provides ultra-low-latency fabric. We follow Nvidia superpod architecture. To cater to high speed read/write, a separate peta-byte scale PFS with 600 GPBS high speed fabric is mounted. We use Slurm for resource management and scheduling, which orchestrates weeks-long training runs by managing node allocation and enabling fault tolerance across the cluster.

\begin{table}[ht]
\centering
\caption{Training costs for CPT and SFT on B200 nodes (calculated based on its rental price, \$6 per GPU-hour).}
\vspace{0.5em}
\label{tab:training-costs-b200}
\begin{tabular}{lcccccc}
\toprule
\textbf{Task} & \textbf{Total GPU-hours} & \textbf{Cost (\$)} \\
\midrule
CPT & 36{,}864 & \$221{,}184 \\
SFT & 768 & \$4{,}608 \\
\bottomrule
\end{tabular}
\\[0.5em]
\textit{Note: Costs for H100 GPUs used for synthetic data generation and ablations are not included.}
\end{table}

\subsubsection{Tokenization}
In this work, we introduce a tokenization strategy that incorporates specialized tokens, including language-aware separators, syntax markers, and structured instruction tokens, to improve parsing precision and enable advanced model behaviors such as Fill-in-the-Middle (FIM) for partial code completion, reasoning-driven approaches for debugging and optimization, and seamless adaptation to Jupyter notebook environments for cell-based execution and iterative development. From an implementation perspective, to maintain compatibility and efficiency, we adopt the Byte-Pair Encoding (BPE) tokenizer from the base model with a 64K token vocabulary, reserving dedicated spaces for specialized tokens. Furthermore, we extend this framework by introducing two reasoning tokens that enhance structured reasoning and context-aware problem solving, thereby improving the model’s capability to handle complex computational tasks effectively.

\subsubsection{Training Hyperparameters}
Consistent with the adopted methodology, we employ \textbf{BF16 mixed precision} to optimize computational efficiency while maintaining numerical stability during training. For attention mechanism, we utilize \textbf{Grouped Query Attention (GQA)}, which improves memory usage and inference speed. The optimization process leverages \textbf{Distributed Fused Adam}, configured with $\beta_{1} = 0.9$, $\beta_{2} = 0.95$, and a weight decay of $0.01$, ensuring robust convergence and regularization. The initial learning rate is set to $8 \times 10^{-5}$, and we apply a warmup strategy with \textbf{5{,}000 steps} to stabilize early training dynamics. These hyper-parameters collectively enhance scalability and performance across large-scale distributed environments.
\subsection{Continual Pre Training} The continual pre-training (CPT) phase focuses on extending the model's capabilities from enterprise text understanding to code intelligence. Starting with a baseline model trained on enterprise text, CPT adapts it for coding tasks through large-scale next-token prediction, by incorporating Fill-in-the-Middle (FIM) for structural infilling. To adapt it for coding tasks, we first trained the model on 50B tokens of code-only data, creating an initial code-focused baseline. This baseline served as the foundation for the CPT stage, upon which subsequent ablation studies were conducted. 

\subsubsection{Data Composition Ablation}
\label{sec:data-ablation}
The data composition ablations were conducted in two stages to determine the optimal distribution of code, natural language text, and mathematical reasoning content on the code-only baseline.

\paragraph{Stage 1 - Ablations on text variability (CPT-S1):} This stage focused on incorporating natural language text into the training mix. Initial reference points for code and text distribution were taken from Qwen2.5-Coder~\cite{qwen2.5} and DeepSeek-Coder-V2~\cite{zhu2024deepseekv2}, and a series of controlled experiments were conducted around these suggested ratios to determine an optimal distribution. The tested configurations explored various code-to-text ratios. Following systematic evaluation, we finalized a distribution that maintains strong code-centric performance while significantly enhancing language understanding capabilities. The resulting ratio is approximately 72--78\% code and 10--15\% text.
 
\paragraph{Stage 2 - Math Integration and Ratio Consolidation (CPT-S2):} Mathematical data was introduced to strengthen logical problem-solving skills, which are essential for coding tasks involving algorithms, numerical computation, and quantitative analysis. Building on the finalized code-text distribution from Stage 1, math data was incorporated and tuned alongside the existing mix to achieve an optimal balance of code,text and math ratio.

\begin{table}[H]

\centering

\caption{Stage 2 - Ablation Study: Math Variability Impact.}

\vspace{0.5em} 

\label{tab:stage2-ablation}

\begin{tabular}{lcccc}

\toprule

\textbf{Data Composition Ratio} & \textbf{HumanEval} & \textbf{HumanEval+} & \textbf{CruxEval-O} & \textbf{PandasEval} \\

\midrule

92.7 : 4.9 : 2.4 & 18.29\% & 14.63\% & 12.38\% & 36.63\% \\

\textbf{78 : 12 : 10 (Finalized Mix)} & \textbf{17.68\%} & \textbf{15.24\%} & \textbf{17.50\%} & \textbf{37.62\%} \\

75.8 : 12.1 : 12.1 & 17.07\% & 14.63\% & 14.88\% & 36.63\% \\

\bottomrule

\end{tabular}

\end{table}

\paragraph{Key Findings:} The CPT ablation studies finalized a balanced data mix of 78\% code, 12\% text, and 10\% math, ensuring strong coding and reasoning capabilities across multiple evaluation benchmarks.
 
\paragraph[Stage 3]{Stage 3 - Scaling the Data for Robustness (CPT-S3):}\label{stage3}

Early stage of evaluations were conducted using 50B tokens for efficiency during testing. Based on the insights gained from these initial runs, we progressively scaled the dataset—starting from 100B tokens, then 150B, and finally 200B—to assess whether performance improvements remained consistent with increased data volume. The experiments confirmed that scaling preserved the performance profile without introducing degradation, reinforcing the robustness of the chosen mix and training strategy.

\paragraph{Key Findings:} Scaling experiments confirmed stability up to 200B tokens without performance degradation, validating the robustness of the finalized data composition and training strategy.
 
\subsubsection{FIM Ratio Ablation}
 
\paragraph{\textbf{Fill-in-the-Middle Objective:}} 

The second experimental objective for our model is Fill-in-the-Middle (FIM), a technique designed to enable structural code completion by predicting missing content using both the preceding and succeeding context. Traditional next-token prediction is insufficient for coder models because programming languages rely on dependencies across different parts of the code. FIM overcomes this limitation by dividing text into three segments, shuffling their order, and linking them with special tokens—effectively creating a fill-in-the-blank objective during pretraining \cite{bavarian2022}.

An exhaustive search for the optimal FIM probability on the larger model trained with code, natural language text and mathematical reasoning content, is computationally prohibitive. To address this, we adopted a more efficient approach by leveraging a smaller coder-only baseline model trained on 50B tokens. This allowed us to establish a reliable reference FIM probability while minimizing time and resource overhead.
 
\paragraph{Stage 4 - FIM Optimization Across Configurations (CPT-S4):}

We started with a code-only baseline and confirmed its effectiveness using a 0.3 FIM ratio. Building on this, we moved to the full configuration and experimented with FIM ratios of 15\% and 30\%. The results showed minimal difference between these two settings, so to strike a balance, we selected an intermediate value and locked the final FIM ratio at 25\%, offering the best trade-off between structural understanding and overall performance.
 
\begin{table}[H]

\centering

\caption{FIM Impact on Performance.}

\vspace{0.5em} 

\label{tab:fim-impact}

\begin{tabular}{lcccc}

\toprule

\textbf{FIM Impact} & \textbf{MBPP} & \textbf{HumanEval} & \textbf{CruxEval-I} & \textbf{CruxEval-O} \\

\midrule

FIM = 15\% & 28.89\% & 15.85\% & 12.13\% & 17.50\% \\

FIM = 30\% & 29.40\% & 15.24\% & 14.25\% & 18.38\% \\

\bottomrule

\end{tabular}

\end{table}

\paragraph{Key Findings:} FIM ratio was optimized from initial trials to 25\% for the final configuration, offering the best trade-off between structural code understanding and overall performance.
 
\subsubsection{Deriving Epoch Strategy}

\paragraph{Stage 5 - Optimal number of epochs (CPT-S5):}

Several experiments were conducted to determine the optimal number of epochs with scaled and the finalized data mix. Different epoch configurations were tested to observe convergence behavior and performance stability. After a thorough evaluation, we determined that three epochs represent the optimal setting, providing the best balance between training efficiency and mitigating overfitting.

\paragraph{Key Findings:} Epoch strategy evaluations determined three epochs as optimal for convergence without overfitting, establishing a robust foundation for subsequent fine-tuning stages.
\subsection{Supervised Fine Tuning}
\label{sec:sft}
Our supervised fine-tuning (SFT) ablations were closely aligned with the progress of continual pre-training. Instead of anchoring experiments to a single pre-training snapshot, we upgraded the SFT runs as stronger checkpoints became available—starting with CPT-S3 (post CPT Stage-3 ablation), then advancing to CPT-E1.0 (1.0-epoch completion), CPT-E1.5 (1.5-epoch completion), and finally CPT-E1.875 (1.875-epoch completion). This rolling integration ensured that the ablations measured the true effects of data composition, scaling, and task mix choices without being confounded by pre-training maturity.
 
\subsubsection{Code Task-specific ablation}
The objective of these experiments was to evaluate the impact of task-specific data subsets on model performance during fine-tuning from the CPT-S3 checkpoint. We applied targeted ablations by filtering high-quality data from a large corpus and fine-tuning the baseline model separately for each task category. This approach allowed us to identify the optimal composition for individual capabilities.
The ablations focused on five key dimensions of coding proficiency: instruction-following, evaluation-style tasks, bug-fixing, reasoning-focused challenges, and function/tool usage. For each dimension, we created fine-tuning mixes emphasizing its representation to observe performance variations across standard benchmarks.\\

\textbf{Key Findings:} Filtering and leveraging high-quality, task-specific subsets for fine-tuning proved significantly more effective than training on the entire dataset, enabling better specialization and overall performance improvements.
\subsubsection{Curated and Synthetic Data Composition and Scaling Ablations}
This stage used CPT-E1.0 checkpoint and was executed in two phases. In the first phase, ablations were performed using the filtered curated data identified from the previous step to determine the optimal configuration for integrating these task-specific datasets effectively. In the second phase, a fresh instance of CPT-E1.0 checkpoint was initialized to apply this configuration while scaling both curated and synthetic datasets from 300k to 800k to 1M samples, ensuring progressive exposure to diverse coding tasks, reasoning patterns, and tool-use scenarios. The best-performing ablation from first phase served as a baseline for validating the effectiveness of this scaling approach.\\

We scaled the dataset progressively, adding complexity at each stage. At 300k samples, the focus was on evaluation-oriented coding tasks to establish a clean baseline for core capabilities. At 800k samples, the configuration introduced a balanced mix of structured code generation, infilling tasks, and curated real-code examples to improve generalization. Finally, at 1M samples, the dataset expanded with repair-oriented samples, multi-step tool-calling scenarios, and large-scale synthetic augmentation to maximize robustness and end-to-end task performance.\\

\textbf{Key Findings:}
Progressive scaling combined with curated and synthetic data integration significantly improved generalization, robustness, and specialized capabilities compared to static configurations.
\begin{table}[H]
\centering
\caption{Benchmark results for different sample sizes.}
\label{tab:benchmark-samples}
\begin{tabular}{lcccccc}
\toprule
Configuration & MBPP & MBPP+ & HumanEval & HumanEval+ & BFCL-v1 & BFCL-v2 \\
\midrule
baseline        & 42.71\% & 38.89\% & 26.22\% & 23.17\% & 24.81\% & 25.58\% \\
baseline+300k   & 53.77\% & 49.74\% & 39.02\% & 35.37\% & 72.19\% & 44.45\% \\
baseline+800k   & 67.09\% & 63.49\% & 43.29\% & 39.63\% & 70.75\% & 44.52\% \\
baseline+1M     & 69.10\% & 63.76\% & 41.46\% & 37.20\% & 70.88\% & 44.71\% \\
\bottomrule
\end{tabular}
\end{table}
\subsubsection{Function-calling Emphasis Ablations}
CPT-E1.5 checkpoint was used in this section to validate refinements and measure their impact. The function-calling ablations focused on evaluating how different data composition strategies influence structured output reliability. The initial phase varied sampling ratios across function-calling–oriented datasets to examine their effects on schema adherence, argument correctness, and tool-selection accuracy, while simultaneously verifying that performance on existing tasks remained stable. After establishing these baselines, a unified JSON schema was introduced to standardize the representation of function-call semantics. Because this refinement modified the underlying data structure, subsequent ablations re-optimized the sampling mixes under the schema-aligned format, again ensuring that improvements in function-calling capability did not come at the expense of previously learned behaviors. During this process, the data was scaled progressively from 800K to 960K and then to 1.1M samples, supporting broader coverage and stability. This enabled a controlled assessment of how representation consistency and data balance jointly shape function-calling robustness without regressing core model competencies.\\

\textbf{Key findings:}
Introducing a unified JSON schema improved the consistency and accuracy of function calls. Rebalancing the data after this change helped maintain overall performance while making function-calling more reliable, ensuring these improvements did not negatively affect other tasks.
\begin{table}[H]
\centering
\label{tab:benchmark-ablations-transposed}
\caption{Performance across different ablations.}
\begin{tabular}{lcccccc}
\toprule
Configuration & MBPP & MBPP+ & HumanEval & HumanEval+ & BFCL-v1 & BFCL-v2 \\
\midrule
CPT-E1.5+800K & 58.79\% & 53.70\% & 35.98\% & 30.49\% & 58.94\% & 38.71\% \\
CPT-E1.5+960K & 59.05\% & 54.23\% & 36.59\% & 32.93\% & 68.75\% & 43.83\% \\
CPT-E1.5+1.1M & 65.58\% & 61.11\% & 37.80\% & 34.15\% & 73.81\% & 59.35\% \\
\bottomrule
\end{tabular}
\end{table}
\subsubsection{Safety Alignment}
Safety was treated as a core training objective rather than as a post-hoc filter. The SFT safety mix combined a calibrated ratio of harmful and general samples with code security exemplars. Through systematic experimentation with varying data composition ratios, we dynamically adjusted the proportion of safety focused content throughout training to achieve optimal detection capabilities while minimizing overcautious behaviors such as excessive refusals on benign inputs. Building on optimal configurations from earlier checkpoints, CPT-E1.875 served as the foundation for safety and robustness experiments. To further evaluate robustness and compliance, RAI datasets were incorporated, and training progress was tracked using global steps to represent extended optimization rather than true epochs.

\paragraph{Takeaways from Ablations:}The ablations collectively highlight how data composition and structured optimization drive measurable improvements in model capability. Task-focused mixes strengthened instruction-following and reasoning, while schema-aligned function-calling enhanced reliability in tool-oriented behaviors. Safety was integrated as a core objective, ensuring compliance without sacrificing utility. Taken together, these results confirm that an iterative, checkpoint-guided process can effectively balance accuracy, structure, and safety, leading to a more robust and dependable code model.
\subsection{Alignment@Scale}

The alignment process was implemented on a custom platform developed by our organization, named \textbf{Alignment@Scale}. This platform enabled the systematic and large-scale integration of human feedback to ensure model outputs remained consistent with predefined objectives. The process began by collecting evaluations from multiple human annotators for each model-generated response. Specifically, three independent annotators assessed candidate responses for a given prompt. This multi-annotator approach provided a more reliable signal of quality and preference, reducing individual bias and enhancing robustness.

Feedback was organized into two key components: \textit{fields} and \textit{questions}. The field denoted the content under evaluation, such as a code-generation prompt or an answer-generation prompt. The question component specified the type of feedback required, which could take one or more formats, including \textit{Label, Multi-label, Ranking, Text, Span, Rating}, and \textit{Code}. Based on manual observation of the response, annotators supplied feedback in multiple question formats, depending on the requirements of the training dataset.

The aggregated feedback was then leveraged to improve the model through \textbf{Reinforcement Learning from Human Feedback (RLHF)}~\cite{ouyang2022traininglanguagemodelsfollow}. By retraining the model on these human-preference signals, its behavior was progressively aligned with desired objectives such as helpfulness, safety, and factual accuracy. This iterative feedback loop enabled alignment at scale, ensuring that as the model’s capabilities expanded, it remained grounded in human values and task-specific goals.
\subsection{Quantization}

For the Mify models, multiple quantization strategies were systematically evaluated to determine their impact on model accuracy and inference performance. The comparative results are presented in Table~\ref{tab:quantization-accuracy}, highlighting the relative accuracy degradation associated with each configuration. Among the evaluated techniques, FP8 quantization was identified as the optimal choice, offering a favorable trade-off between computational efficiency and accuracy preservation.
\begin{table}[H]
\centering
\caption{Accuracy change across different quantization configurations.}
\label{tab:quantization-accuracy}
\begin{tabular}{|l|l|}
\hline
\textbf{Quantization} & \textbf{Accuracy change \%} \\
\hline
FP16 & baseline \\
8-bit SmoothQuant & -0.37 \\
8-bit AWQ & -1.99 \\
FP8 & -1.35 \\
\hline
\end{tabular}
\end{table}
The FP8 quantization configuration was selected as the preferred approach due to its superior throughput characteristics and significantly reduced token generation latency, while incurring only negligible accuracy degradation.

\paragraph{Throughput Analysis}
Our FP8-quantized model was optimized using the \textbf{TensorRT-LLM engine}\footnote{\url{https://github.com/NVIDIA/TensorRT-LLM}}. The achieved throughput across hardware platforms is summarized in Table~\ref{tab:hardware-throughput}. H100 delivers an order-of-magnitude higher throughput than other GPU classes, indicating substantially faster inference on H100 for the FP8 quantized TensorRT-LLM configuration used in this work.

\begin{table}[H]
\centering
\caption{Token generation throughput across hardware platforms.}
\label{tab:hardware-throughput}
\begin{tabular}{|l|l|}
\hline
\textbf{Hardware} & \textbf{Output tokens per second} \\
\hline
L40S & 885 \\
A100 & 1047 \\
H100 & 8658 \\
\hline
\end{tabular}
\end{table}

FP8 quantization offers a balanced solution for optimizing large language model deployment, combining high throughput and reduced latency with acceptable accuracy trade-offs. Together with TensorRT-LLM optimizations and GGUF-based CPU inference, FP8 enables scalable and efficient deployment across diverse hardware environments.


\section{Evaluation}

We evaluate our model, \textbf{Mify-Coder}, alongside several open-source models across a range of benchmarks. The evaluation includes various benchmarks like HumanEval, HumanEval+, MBPP, MBPP+, BFCL-v2 covering tasks like code generation and function calling scenarios.

\subsection{Tasks}

\subsubsection{Code Generation}

These benchmarks test a model’s ability to generate correct Python code from text. HumanEval \cite{chen2021evaluating} and HumanEval+ \cite{liu2023your} focus on writing Python functions, with HumanEval+ adding stricter tests. MBPP \cite{austin2021program} and MBPP+ check algorithmic reasoning, covering more cases.

\subsubsection{Function Calling}

This benchmark test the model’s capability to produce structured function calls from natural language inputs. BFCL-v2 \cite{yan2024berkeley} introduces more complex signatures and stricter evaluation, emphasizing precision and consistency in structured outputs.

\subsection{Evaluation Setup}

\subsubsection{Baselines}

We compare our proposed model against a diverse set of models, including Claude Sonnet 4.5, DeepSeek-Coder-33B-Instruct\cite{guo2024deepseek}, Qwen3-Coder-30B-A3B-Instruct\cite{yang2025qwen3}, Llama-3.3-70B-Instruct, OLMo-3-7B-Instruct, Mistral-Large-2407-v1, GPT-4o\cite{hurst2024gpt}. All models were evaluated using publicly available checkpoints or official APIs, ensuring a fair and consistent comparison across both code generation and function-calling tasks. This setup highlights the relative strengths of models trained with different strategies. 
\subsubsection{Decoding Strategy}

For all models evaluated, including our proposed \textbf{Mify-Coder}, we adopt a consistent decoding strategy to ensure fair comparison across tasks. We have used  pass@1 as metric and specifically, the maximum generation length (\texttt{max\_gen\_tokens}) is varied per benchmark to better align with task-specific requirements. This configuration balances precision and output diversity and is consistently applied across both pretrained and instruction-tuned models. The benchmark-specific \texttt{max\_gen\_tokens} settings are as follows: MBPP (1024), MBPP+ (1024), BFCL-v2 (1024), HumanEval (1024), HumanEval+ (1024).

\subsection{Results of Code Generation}
\begingroup
\setlength{\intextsep}{-1pt}   
\setlength{\textfloatsep}{-1pt}
\begin{table}[H]
    \centering
    \footnotesize
    \caption{Code Generation Results on Benchmarks.}
    \begin{tabular}{l c c c c c}
        \toprule
        \textbf{Model Name} & \textbf{Model Size} & \textbf{MBPP} & \textbf{MBPP+} & \textbf{HumanEval} & \textbf{HumanEval+} \\
        \midrule
        \multicolumn{6}{c}{\textbf{Small Models ($<$ 10B)}} \\
        \midrule
        OLMo-3-7B-Instruct & 7B & *45.73\% & 60.20\% & *54.27\% & 77.20\% \\
        Mify-Coder & 2.5B & *91.21\% & *89.15\% & *53.66\% & *48.78\% \\
        \midrule
        \multicolumn{6}{c}{\textbf{Medium Models (10B–50B)}} \\
        \midrule
        Qwen3-Coder-30B-A3B-Instruct & 30B & 87.60\% & 73.50\% & 92.10\% & 87.80\% \\
        DeepSeek-Coder-33B-Instruct & 33B & 70.00\% & 70.10\% & 79.30\% & 75.00\% \\
        \midrule
        \multicolumn{6}{c}{\textbf{Large Models ($\geq$ 50B)}} \\
        \midrule
        Mistral-Large-2407-v1 & 123B & 80.00\% & 69.00\% & 92.00\% & 87.00\% \\
        Llama-3.3-70B-Instruct & 70B & 87.60\% & *71.96\% & 88.40\% & *42.68\% \\
        \midrule
        \multicolumn{6}{c}{\textbf{Closed-APIs}} \\
        \midrule
        Claude Sonnet 4.5 & -- & 88.70\% & *64.81\% & 92.30\% & *90.24\% \\
        GPT-4o & -- & 87.60\% & *53.17\% & 90.02\% & 87.20\% \\
        \bottomrule
    \end{tabular}
\end{table}

Where applicable, reported benchmark scores were prioritised for inclusion. In instances where published results were unavailable, performance metrics were derived through local execution; scores obtained via empirical execution are denoted with an asterisk (*). For the MBPP and MBPP+ benchmarks, the reported values represent the maximum performance observed between the 0-shot and 3-shot execution configurations. Consequently, while MBPP and HumanEval scores typically exceed those of their "plus" counterparts (MBPP+ and HumanEval+), this trend is not observed for the OLMo-3-7B-Instruct model due to the specific selection criteria applied to reported versus executed results.

Evaluation on standard code-generation benchmarks indicates that mid-sized models (10B-50B), including Qwen3-Coder-30B-A3B-Instruct and DeepSeek-Coder-33B-Instruct, achieve strong overall performance. Qwen3-Coder-30B-A3B-Instruct attains 87.60\% on MBPP and 92.10\% on HumanEval, whereas DeepSeek-Coder-33B-Instruct records 70.00\% on MBPP and 79.30\% on HumanEval. Among large-scale models (>50B), Meta-Llama-3.3-70B-Instruct achieves 87.60\% on MBPP and 88.40\% on HumanEval; however, performance decreases substantially on HumanEval+ (42.68\%). In contrast, smaller models (<10B) such as OLMo-3-7B-Instruct yield lower scores (45.73\% on MBPP and 54.27\% on HumanEval). Notably, Mify-Coder (2.5B) achieves 91.21\% on MBPP and 89.15\% on MBPP+, representing the best performance among the evaluated models on MBPP related tasks, while remaining competitive on HumanEval (53.66\%) and HumanEval+ (48.78\%). Overall, these results suggest that Mify-Coder offers a favorable accuracy–efficiency balance relative to significantly larger models.
\subsection{Results of Function Calling}

\begin{table}[H]
    \centering
    \footnotesize
    \caption{Function Calling Results on BFCL-v2.}
    \begin{tabular}{l c c}
          \toprule
          \textbf{Model Name} & \textbf{Model Size} & \textbf{BFCL-v2} \\
        \midrule
        \multicolumn{3}{c}{\textbf{Small Models ($<$ 10B)}} \\
        \midrule
        OLMo-3-7B-Instruct & 7B & *12.03\% \\
        Mify-Coder & 2.5B & *55.14\% \\
        \midrule
        \multicolumn{3}{c}{\textbf{Medium Models (10B–50B)}} \\
        \midrule
        Qwen3-Coder-30B-A3B-Instruct & 30B & *76.63\% \\
        DeepSeek-Coder-33B-Instruct & 33B & 45.40\% \\
        \midrule
        \multicolumn{3}{c}{\textbf{Large Models ($\geq$ 50B)}} \\
        \midrule
        Mistral-Large-2407-v1 & 123B & *71.12\% \\
        Llama-3.3-70B-Instruct & 70B & 77.30\% \\
        \midrule
        \multicolumn{3}{c}{\textbf{Closed-APIs}} \\
        \midrule
        Claude Sonnet 4.5 & -- & *74.30\% \\
        GPT-4o & -- & *71.70\% \\
        \bottomrule
    \end{tabular}
\end{table}

On the \textbf{BFCL-v2} benchmark, which evaluates tool-use and reasoning capabilities, \textbf{Mify-Coder (2.5B)} achieves an accuracy of \textbf{55.14\%}, demonstrating competitive performance despite its compact size. Larger models such as \textbf{Qwen3-Coder-30B-A3B-Instruct} reach higher scores (\textbf{76.63\%}), while \textbf{Llama-3.3-70B-Instruct} delivers \textbf{77.30\%}. In contrast, \textbf{DeepSeek-coder-33B-Instruct} records a moderate \textbf{45.40\%}, and smaller models like \textbf{OLMo-3-7B-Instruct} achieve only \textbf{12.03\%}. These results highlight that smaller, well-optimized models such as Mify-Coder can deliver strong tool-use performance without the computational overhead associated with significantly larger architectures.
\begin{figure}[htbp]
    \centering
    \includegraphics[width=1\textwidth]{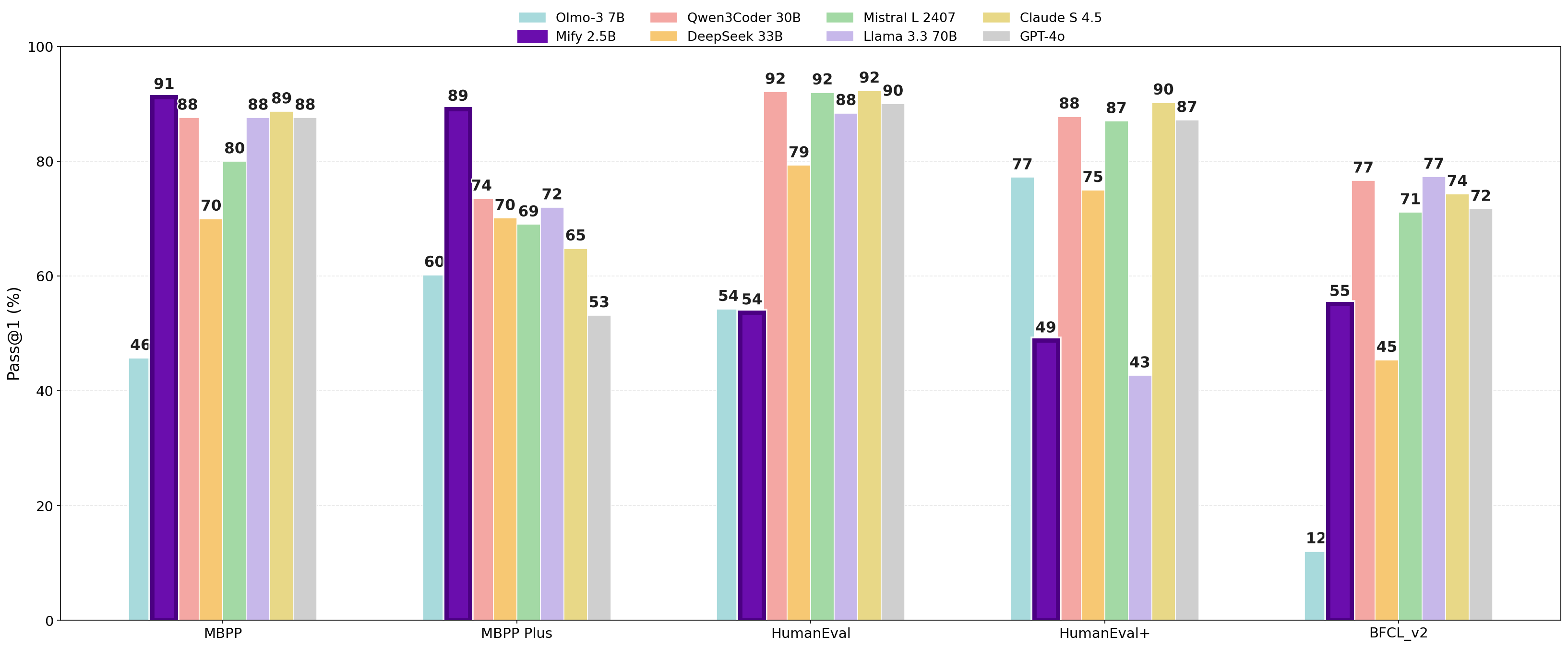}
    \caption{The code generation, Function calling performance of Coder Models on seven benchmarks MBPP, MBPP+, HumanEval, HumanEval+, BFCL-v2.}
\end{figure}

\subsection{Low-Latency Optimization and Throughput Results}
\noindent The model was meticulously optimized for \textbf{ultra-low latency}, achieving an impressive \textbf{time-to-first-token (TTFT) of just 0.10 seconds}, which enables near-instantaneous response initiation for real-time applications such as interactive chatbots and streaming generation. Furthermore, it delivers a remarkable \textbf{throughput of 6{,}645.3 tokens per second} on a single \textbf{NVIDIA H100 (80GB) GPU}, ensuring high-volume text generation without compromising efficiency. These performance metrics significantly outperform most \textbf{sub-8B small language models (SLMs)} in both latency and generation speed, making the model highly competitive in production environments. This level of optimization was achieved through a combination of \textbf{kernel-level acceleration}, \textbf{fused attention mechanisms}, and \textbf{memory footprint reduction techniques}, which collectively minimize computational overhead and maximize GPU utilization. As a result, the model is ideally suited for \textbf{low-latency inference scenarios}, \textbf{enterprise-scale deployments}, and \textbf{edge computing environments} where responsiveness, scalability, and cost-efficiency are critical.\subsection{Performance Evaluation of Quantized Configurations}
\noindent Furthermore, the quantized versions of Mify-Coder enable seamless deployment on standard desktop environments, eliminating the need for specialized hardware such as GPUs. The accompanying chart compares accuracy across different quantized configurations on benchmarks for code generation (HumanEval, HumanEval+) and tool use (BFCL-v2). While reducing memory footprint significantly—from 4.67 GB to as low as 1.43 GB—the models maintain competitive accuracy. These results demonstrate that quantization offers substantial efficiency gains without severely compromising performance, making the models practical for resource-constrained environments.
\begin{figure}[H]
    \centering
    \includegraphics[width=0.8\textwidth]{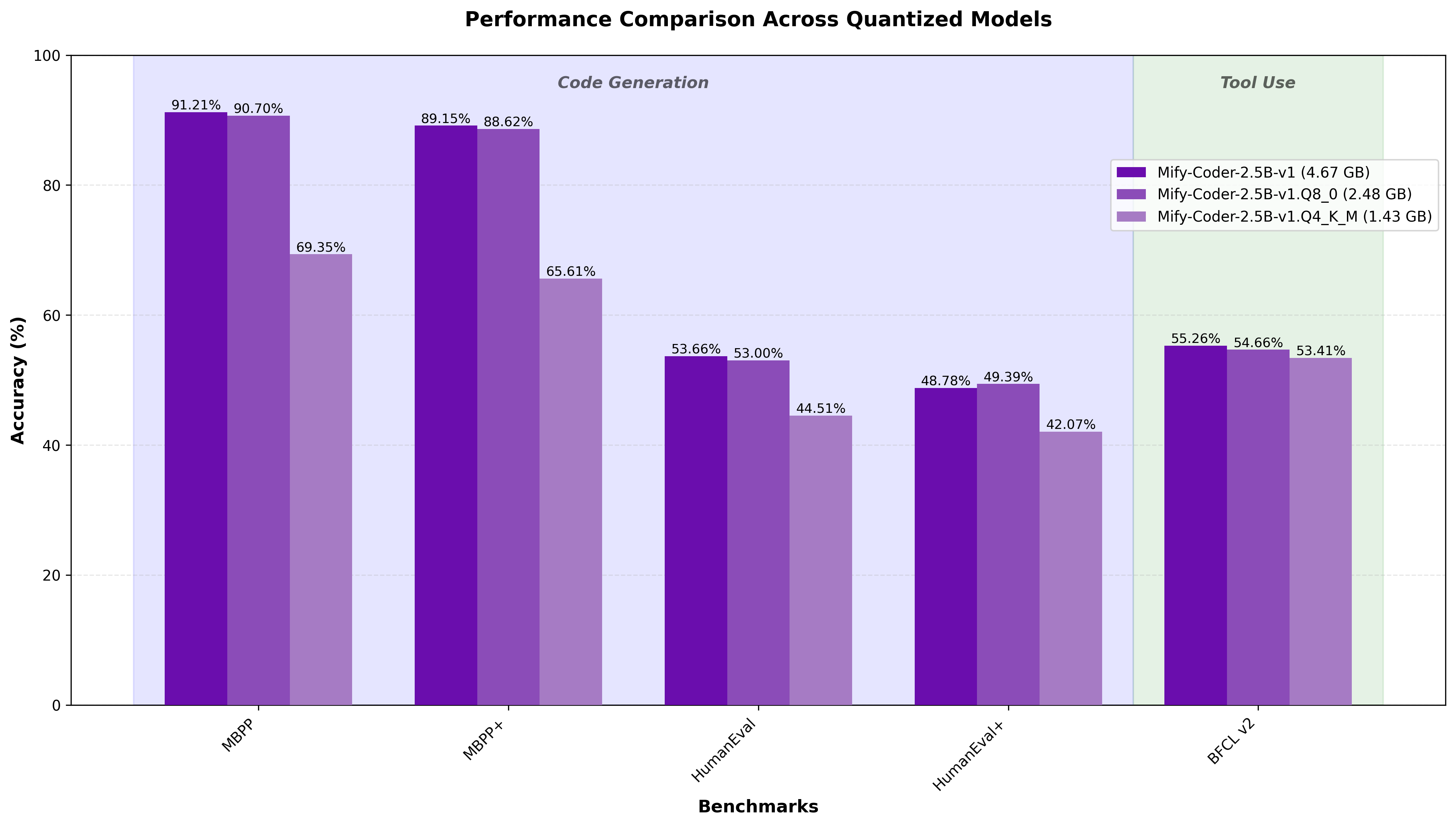}
    \caption{Accuracy comparison across quantized Mify-Coder configurations on code generation and tool-use benchmarks.}
\end{figure}    
\noindent

\section{Responsible AI}
Developing AI systems that are safe, harmless, and helpful requires a principled and systematic approach to safety training. Our methodology ensures that Mify-Coder mitigates harmful outputs while preserving contextual accuracy, technical precision, and relevance in both natural language and code generation tasks.
 
\noindent The approach is grounded in two core principles: Harmlessness, which focuses on preventing responses with toxic language and insecure code that could introduce vulnerabilities or ethical risks, and Helpfulness, which ensures responses remain accurate, relevant, and user centric. Together, these principles balance safety with functionality and guide our dataset design, training methodology, and evaluation framework. To achieve this, we curated datasets combining high-quality open-source repositories with synthetic samples, addressing two dimensions: Textual Safety for ethical language generation and Secure code generation for best coding practices. 

\noindent In the area of text safety, our primary goal was to enhance the model’s ability to detect and mitigate harmful language while maintaining fluency and helpfulness. To accomplish this, we performed a series of supervised fine-tuning experiments using datasets that systematically varied the ratio of harmful to general samples. Early experiments with a higher proportion of harmful content improved unsafe pattern detection but led to increased refusal rates and false positives on benign prompts, indicating an overemphasis on safety. By iteratively adjusting the mixing ratio by gradually reducing harmful samples while preserving a strong baseline of general data, we identified an optimal balance of 1:4 ratio between harmful and general samples. This approach enabled the model to robustly identify genuinely unsafe content without sacrificing linguistic quality or user-centric helpfulness. As a result, our methodology achieves a significant reduction in harmful outputs while preserving natural language fluency, as validated by Stanford's AIR-Bench safety benchmark.

\noindent In the domain of code generation, our primary objective is to strengthen secure code generation capabilities while preserving overall utility and helpfulness. To achieve this, we conducted extensive experiments using datasets comprising vulnerable code samples paired with their secure counterparts, alongside general-purpose utility code. Remarkably, we observed substantial improvements through supervised fine-tuning on a large-scale dataset that maintained a balanced mix of utility and security-focused code. This approach enabled us to achieve performance on par with leading language models globally for secure code generation, as validated by Meta’s CyberSecEval autocomplete benchmark.
 
\noindent After Supervised Fine-Tuning, we benchmarked performance using industry-standard safety metrics and custom evaluations for harmfulness detection and insecure code identification. Results demonstrate a significant reduction in harmful outputs across text and achieve secure code generation, alongside improved helpfulness scores, ensuring responses remain actionable and user centric. The heatmaps \textit{Figure}\ref{fig:AirBench}, \textit{Figure}\ref{fig:CyberSecEval4} below illustrate performance comparisons of frontier models and our model across Stanford-AIR-Bench \cite{airbench2024} and CyberSecEval4 \cite{cyberseceval2024} benchmarks respectively. 
\begin{figure}[H]
    \centering
    \includegraphics[width=0.8\textwidth]{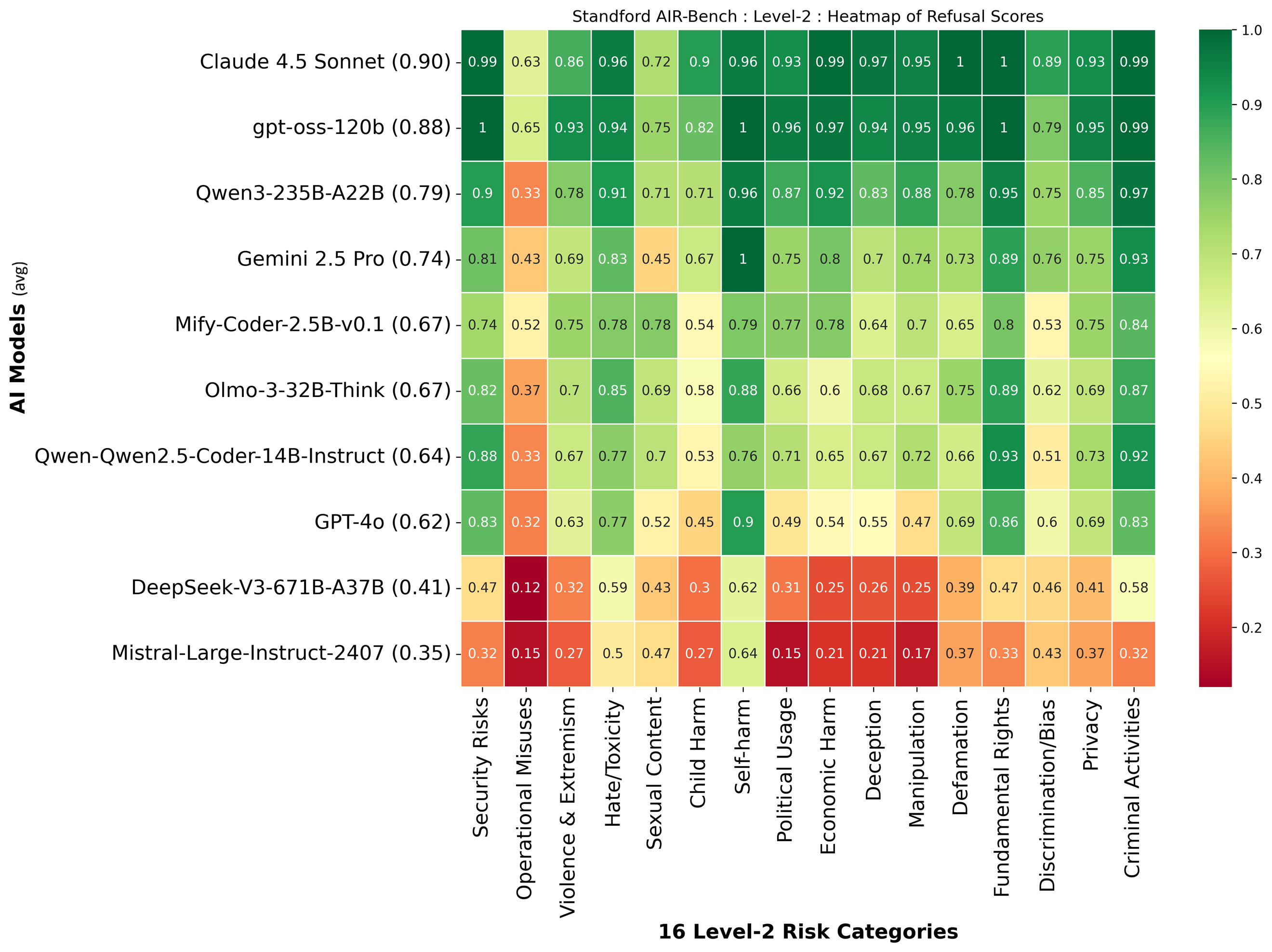}
    \caption{Standford AIR-Bench-Safety Assessment Across Jurisdiction. }
    \label{fig:AirBench}
\end{figure}

\begin{figure}[H]
    \centering
    \includegraphics[width=0.8\textwidth]{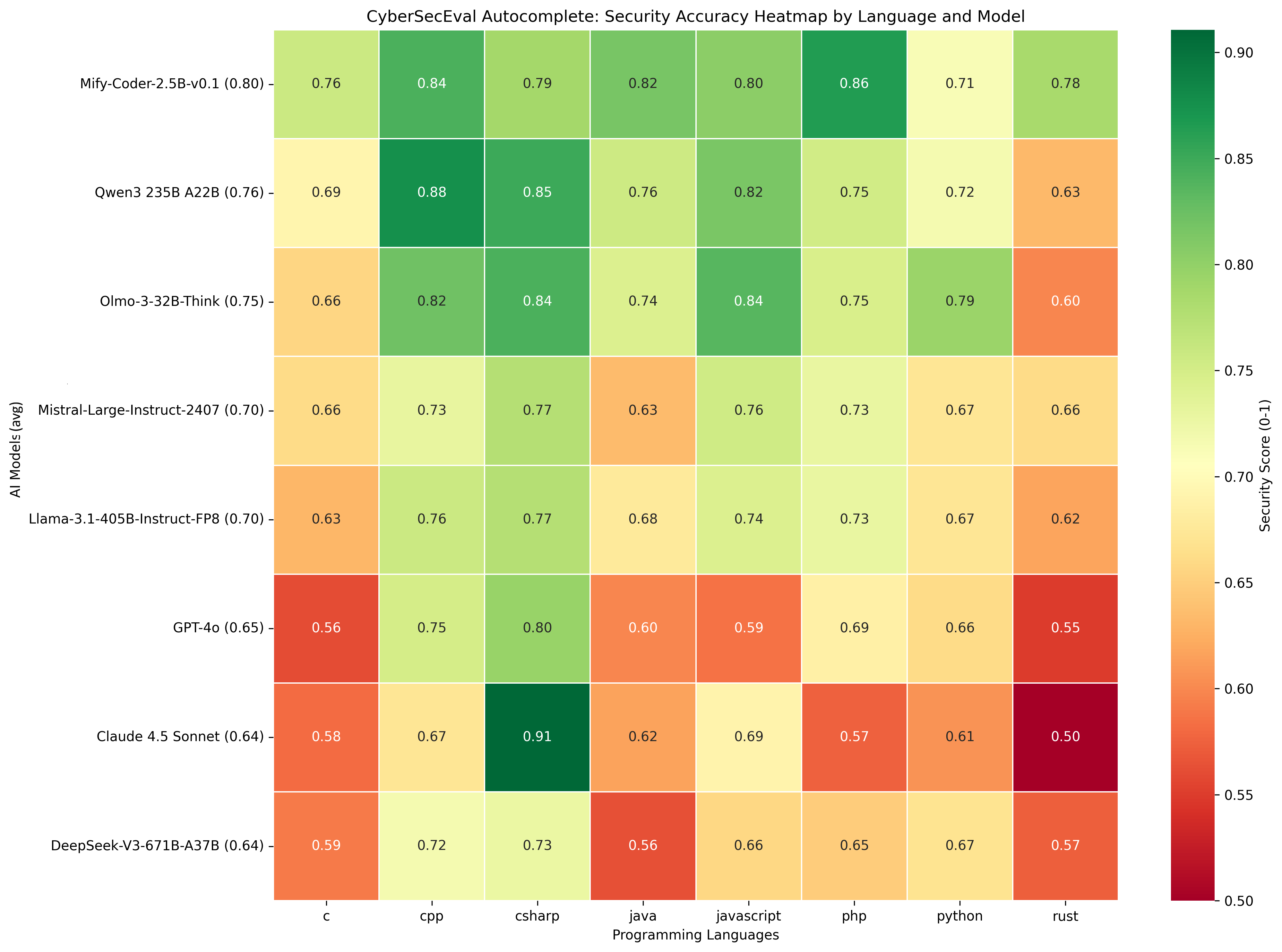}
    \caption{Secure code generation capabilities of language models based on CyberSecEval-Autocomplete. }
    \label{fig:CyberSecEval4}
\end{figure}
\noindent In summary, aligning harmlessness and helpfulness as core principles, maintaining a balanced data ratio, and applying SFT with probability-based sampling collectively deliver measurable improvements in safety and utility. These results affirm the value of structured strategies for Responsible AI.
\section{Conclusion}
Mify-Coder, a 2.5B-parameter model, demonstrates that smaller architectures can achieve frontier-grade performance through compute-efficient strategies and principled data curation. The model integrates continual pretraining and supervised fine-tuning within a disciplined pipeline, leveraging a balanced data mix of 78\% code, 12\% text, and 10\% math finalized via ablation studies. Synthetic data generation plays a pivotal role, employing seed-grounded and seedless strategies with prompt rotation to maximize diversity, followed by LLM-as-a-judge validation for correctness and formatting. Post-processing includes semantic deduplication and benchmark decontamination to ensure integrity. Scaling experiments explored synthetic volumes from 3B to 15B tokens, confirming that synthetic augmentation significantly boosts performance on benchmarks like MBPP and BFCL, especially for code generation and function-calling tasks. Additional ablations optimized FIM ratio, epoch strategy, and data scaling for robustness. Despite its compact size, Mify-Coder achieves exceptional results on MBPP, HumanEval, and strong function-calling scores, while maintaining enterprise-grade safety and enabling CPU-based deployment through quantization. These findings challenge scale-centric assumptions, proving that synthetic augmentation, rigorous quality control, and compute discipline can deliver high-performing, resource-efficient models for real-world coding tasks.
\section{Future Work}

Our future work will address current limitations in complex algorithmic reasoning and long-horizon planning through architectural refinements and advanced training methodologies. These enhancements will enable multi-step decomposition and improved reasoning capabilities, allowing the model to handle increasingly sophisticated computational tasks.

We will extend the model beyond text-only operation by incorporating multi-modal encoders, enabling vision-guided code generation for tasks such as UI-to-code translation and diagram interpretation. This integration will broaden the model’s applicability to real-world development scenarios where visual and textual inputs coexist.

Robustness and accuracy on practical software engineering benchmarks will be strengthened to ensure reliability in production environments. We are looking to evolve the alignment strategy toward reinforcement learning with verifiable rewards (RLVR), leveraging automated test suites and static analysis to guarantee correctness, safety, and compliance with coding standards. Collectively, these research directions will result in a more capable, secure, and versatile system designed to support next-generation software development workflows across diverse modalities and complex reasoning tasks.

\newpage
\appendix
\section{Contributors}
\begin{multicols}{3}
\raggedright
Abhinav Parmar \\
Abhisek Panigrahi \\
Abhishek Kumar Dwivedi \\
Abhishek Bhattacharya \\
Adarsh Ramachandra \\
Aditya Choudhary \\
Aditya Garg \\
Aditya Raj \\
Alankrit Bhatt \\
Alpesh Yadav \\
Anant Vishnu \\
Ananthu Pillai \\
Ankush Kumar \\
Aryan Patnaik \\
Aswatha Narayanan S \\
Avanish Raj Singh \\
Bhavya Shree Gadda \\
Brijesh Pankajbhai Kachhadiya \\
Buggala Jahnavi \\
Chidurala Nithin Krishna \\
Chintan Shah \\
Chunduru Akshaya \\
Debarshi Banerjee \\
Debrup Dey \\
Deepa R. \\
Deepika B G \\
Faiz ur Rahman \\
Gagan Gayari \\
Gudhi Jagadeesh Kumar Naidu \\
Gursimar Singh \\
Harshal Tyagi \\
Harshini K \\
James Mani Vathalloor \\
Jayarama Nettar \\
Jayashree Gajjam \\
Joe Walter Sugil George \\
Kamalakara Sri Krishna Tadepalli \\
Kamalkumar Rathinasamy \\
Karan Chaurasia \\
Karthikeyan S \\
Kashish Arora \\
Kaushal Desai \\
Khushboo Buwade \\
Kiran Manjrekar \\
Malikireddy Venkata Sai Likhitha \\
Manjunath A \\
Mitali Mahavir Bedmutha \\
Mohammed Rafee Tarafdar \\
Nikhil Tiwari \\
Nikitha K Gigi \\
Pavan Ravikumar \\
Pendyala Swarnanjali \\
Piyush Anand \\
Prakash Chandrasekar \\
Prasanna Bhalchandra Gawade \\
Prasanth Sivan \\
Preeti Khurana \\
Priyanshi Babbar \\
Rajab Ali Mondal \\
Rajesh Kumar Vissapragada \\
Rajeshwari Ganesan \\
Rajeswari Koppisetti \\
Ramjee R. \\
Ramkumar Thiruppathisamy \\
Rani G. S. \\
S Reka \\
Samarth Gupta \\
Sandeep Reddy Kothakota \\
Sarathy K \\
Sathyanarayana Sampath Kumar \\
Saurabh Kumar \\
Shashank Khasare \\
Shenbaga Devi Venkatesh Kumar \\
Shiva Rama Krishna Parvatham \\
Shoeb Shaikh \\
Shrishanmathi A \\
Shubham Pathak \\
Sree Samhita Koppaka \\
Sreenivasa Raghavan K S \\
Sreeram Venkatasubramanian \\
Suprabha Desai Bojja \\
Swetha R \\
Syed Ahmed \\
Chinmai Harshitha Thota \\
Tushar Yadav \\
Veeravelly Kusumitha \\
V V S S Prasanth Patnaik\\
Vidya Sri Sesetti \\
Vijayakeerthi K \\
Vikram Raj Bakshi \\
Vinay K K \\
Vinoth Kumar Loganathan \\
Vipin Tiwari \\
Vivek Kumar Shrivastav \\
V Venkata Sri Datta Charan \\
Wasim Akhtar Khan \\
\end{multicols}

\newpage
\bibliographystyle{plain}  
\bibliography{references} 

@article{abouelenin2025phi4mini,
  title={Phi-4-mini Technical Report: Compact yet Powerful Multimodal Language Models via Mixture-of-Loras},
  author={Abouelenin, Abdelrahman and Ashfaq, Atabak and Atkinson, Adam and Awadalla, Hany and Bach, Nguyen and Bao, Jianmin and Benhaim, Alon and Cai, Martin and Chaudhary, Vishrav and Chen, Congcong and others},
  journal={arXiv preprint arXiv:2503.01743},
  year={2025}
}

@article{allal2023santacoder,
  title={SantaCoder: Don't Reach for the Stars!},
  author={Allal, Loubna Ben and Li, Raymond and Kocetkov, Denis and Mou, Chenghao and Akiki, Christopher and Ferrandis, Carlos Munoz and Muennighoff, Niklas and Mishra, Mayank and Gu, Alex and Dey, Manan and others},
  journal={arXiv preprint arXiv:2301.03988},
  year={2023}
}

@article{austin2021program,
  title={Program Synthesis with Large Language Models},
  author={Austin, Jacob and Odena, Augustus and Nye, Maxwell and Bosma, Maarten and Michalewski, Henryk and Dohan, David and Jiang, Ellen and Cai, Carrie and Terry, Michael and Le, Quoc and others},
  journal={arXiv preprint arXiv:2108.07732},
  year={2021}
}

@article{bavarian2022,
  title={Efficient Training of Language Models to Fill in the Middle},
  author={Bavarian, Mohammad and Jun, Heewoo and Tezak, Nikolas and Schulman, John and McLeavey, Christine and Tworek, Jerry and Chen, Mark},
  journal={arXiv preprint arXiv:2207.14255},
  year={2022}
}

@article{chen2021evaluating,
  title={Evaluating Large Language Models Trained on Code},
  author={Chen, Mark},
  journal={arXiv preprint arXiv:2107.03374},
  year={2021}
}

@article{guo2024deepseek,
  title={DeepSeek-Coder: When the Large Language Model Meets Programming--The Rise of Code Intelligence},
  author={Guo, Daya and Zhu, Qihao and Yang, Dejian and Xie, Zhenda and Dong, Kai and Zhang, Wentao and Chen, Guanting and Bi, Xiao and Wu, Yu and Li, YK and others},
  journal={arXiv preprint arXiv:2401.14196},
  year={2024}
}

@article{qwen2.5,
  title={Qwen2.5-Coder Technical Report},
  author={Hui, Binyuan and Yang, Jian and Cui, Zeyu and Yang, Jiaxi and others},
  journal={arXiv preprint arXiv:2409.12186},
  year={2024}
}

@article{muennighoff2023octopack,
  title={OctoPack: Instruction Tuning Code Large Language Models},
  author={Muennighoff, Niklas and Liu, Qian and Zebaze, Armel and others},
  journal={arXiv preprint arXiv:2308.07124},
  year={2023}
}

@article{wang2021codet5,
  title={CodeT5: Identifier-aware Unified Pre-trained Encoder-Decoder Models for Code Understanding and Generation},
  author={Wang, Yue and Wang, Weishi and Joty, Shafiq and Hoi, Steven CH},
  journal={Proceedings of the 2021 Conference on Empirical Methods in Natural Language Processing (EMNLP)},
  year={2021}
}

@misc{xu2022polycoder,
  title={A Systematic Evaluation of Large Language Models of Code},
  author={Xu, Frank F. and Alon, Uri and Neubig, Graham and Hellendoorn, Vincent Josua},
  year={2022}
}

@misc{yan2024berkeley,
  title={Berkeley Function Calling Leaderboard},
  author={Yan, Fanjia and Mao, Huanzhi and Ji, Charlie Cheng-Jie and Zhang, Tianjun and Patil, Shishir G and Stoica, Ion and Gonzalez, Joseph E},
  year={2024}
}

@article{yang2025qwen3,
  title={Qwen3 Technical Report},
  author={Yang, An and Li, Anfeng and Yang, Baosong and Zhang, Beichen and Hui, Binyuan and Zheng, Bo and Yu, Bowen and Gao, Chang and Huang, Chengen and Lv, Chenxu and others},
  journal={arXiv preprint arXiv:2505.09388},
  year={2025}
}

@article{yu2023wavecoder,
  title={WaveCoder: Widespread and Versatile Enhanced Instruction Tuning with Refined Data Generation},
  author={Yu, Zhaojian and Zhang, Xin and Shang, Ning and others},
  journal={arXiv preprint arXiv:2312.14187},
  year={2023}
}

@article{zhu2024deepseekv2,
  title={DeepSeek-Coder-V2: Breaking the Barrier of Closed-Source Models in Code Intelligence},
  author={Zhu, Qihao and Guo, Daya and Shao, Zhihong and Yang, Dejian and others},
  journal={arXiv preprint arXiv:2406.11931},
  year={2024}
}

@article{wei2024arctic,
  title        = {Arctic-SnowCoder: Demystifying High-Quality Data in Code Pretraining},
  author       = {Wei, Yuxiang and Han, Hojae and Samdani, Rajhans},
  journal      = {arXiv preprint arXiv:2409.02326},
  year         = {2024},
  url          = {https://arxiv.org/abs/2409.02326},
  doi          = {10.48550/arXiv.2409.02326}
}

@article{wei2024magicoder,
  title        = {Magicoder: Empowering Code Generation with OSS-Instruct},
  author       = {Wei, Yuxiang and Wang, Zhe and Liu, Jiawei and Ding, Yifeng and Zhang, Lingming},
  journal      = {arXiv preprint arXiv:2312.02120},
  year         = {2024},
  note         = {ICML 2024},
  url          = {https://arxiv.org/abs/2312.02120},
  doi          = {10.48550/arXiv.2312.02120}
}

@article{ahmad2025opencodeinstruct,
  title        = {OpenCodeInstruct: A Large-scale Instruction Tuning Dataset for Code LLMs},
  author       = {Ahmad, Wasi Uddin and Ficek, Aleksander and Samadi, Mehrzad and Huang, Jocelyn and Noroozi, Vahid and Majumdar, Somshubra and Ginsburg, Boris},
  journal      = {arXiv preprint arXiv:2504.04030},
  year         = {2025},
  url          = {https://arxiv.org/abs/2504.04030},
  doi          = {10.48550/arXiv.2504.04030}
}

@misc{ouyang2022traininglanguagemodelsfollow,
      title={Training language models to follow instructions with human feedback}, 
      author={Long Ouyang and Jeff Wu and Xu Jiang and Diogo Almeida and Carroll L. Wainwright and Pamela Mishkin and Chong Zhang and Sandhini Agarwal and Katarina Slama and Alex Ray and John Schulman and Jacob Hilton and Fraser Kelton and Luke Miller and Maddie Simens and Amanda Askell and Peter Welinder and Paul Christiano and Jan Leike and Ryan Lowe},
      year={2022},
      eprint={2203.02155},
      archivePrefix={arXiv},
      primaryClass={cs.CL},
      url={https://arxiv.org/abs/2203.02155}, 
}

@article{hurst2024gpt,
  title={Gpt-4o system card},
  author={Hurst, Aaron and Lerer, Adam and Goucher, Adam P and Perelman, Adam and Ramesh, Aditya and Clark, Aidan and Ostrow, AJ and Welihinda, Akila and Hayes, Alan and Radford, Alec and others},
  journal={arXiv preprint arXiv:2410.21276},
  year={2024}
}

@online{airbench2024,
  author       = {Yi Zeng and Yu Yang and Andy Zhou and Jeffrey Ziwei Tan and Yuheng Tu and Yifan Mai and Kevin Klyman and Minzhou Pan and Ruoxi Jia and Dawn Song and Percy Liang and Bo Li},
  title        = {AIR‑Bench 2024: A Safety Benchmark Based on Risk Categories from Regulations and Policies},
  year         = {2024},
  month        = aug # "~5",
  url          = {https://arxiv.org/abs/2407.17436},
  organization = {arXiv},
}

@online{cyberseceval2024,
  author       = {Shengye Wan and Cyrus Nikolaidis and Daniel Song and David Molnar and James Crnkovich and Jayson Grace and Manish Bhatt and Sahana Chennabasappa and Spencer Whitman and Stephanie Ding and Vlad Ionescu and Yue Li and Joshua Saxe},
  title        = {CyberSecEval 3: Advancing the Evaluation of Cybersecurity Risks and Capabilities in Large Language Models},
  year         = {2024},
  month        = aug # "~2",
  version      = {v1},
  url          = {https://arxiv.org/abs/2408.01605},
  organization = {arXiv},
}

@article{liu2023your,
  title={Is your code generated by chatgpt really correct? rigorous evaluation of large language models for code generation},
  author={Liu, Jiawei and Xia, Chunqiu Steven and Wang, Yuyao and Zhang, Lingming},
  journal={Advances in Neural Information Processing Systems},
  volume={36},
  pages={21558--21572},
  year={2023}
}

@article{lozhkov2024starcoder2,
  title        = {StarCoder 2 and The Stack v2: The Next Generation},
  author       = {Lozhkov, Anton and Li, Raymond and Ben Allal, Loubna and Cassano, Federico and others},
  journal      = {arXiv preprint arXiv:2402.19173},
  year         = {2024},
  url          = {https://arxiv.org/abs/2402.19173},
  doi          = {10.48550/arXiv.2402.19173}
}

\end{document}